\documentclass[10pt,conference,final,oneside,twocolumn,nofonttune]{IEEEtran}
\usepackage{amscd}
\usepackage{amsfonts}
\usepackage{mathrsfs}

\usepackage{graphicx}
\usepackage{cite}
\usepackage{citesort}
\usepackage{color}
\usepackage{psfrag}
\usepackage{subfigure}
\usepackage{amssymb}
\usepackage{epsfig}
\usepackage{pifont}
\usepackage{amsmath}
\usepackage{array}
\usepackage{multicol}
\usepackage{algorithm}
\usepackage{algorithmic}
\allowdisplaybreaks[1]

\renewcommand{\baselinestretch}{0.91}
\newtheorem{prop}{Proposition}
\makeatletter
\def\ifundefined{\@ifundefined}
\makeatother 

\begin{document}

\title{Fast Adaptive S-ALOHA Scheme for Event-driven Machine-to-Machine Communications}

%\author{
%\authorblockN{}
%\authorblockA{} \and
%\authorblockN{}
%\authorblockA{} }

\author{\IEEEauthorblockN{Huasen Wu\IEEEauthorrefmark{1}\IEEEauthorrefmark{2},
Chenxi Zhu\IEEEauthorrefmark{3}, Richard J. La\IEEEauthorrefmark{4},
Xin Liu\IEEEauthorrefmark{2}, and Youguang
Zhang\IEEEauthorrefmark{1}}
\IEEEauthorblockA{\IEEEauthorrefmark{1}School of Electronic and
Information Engineering, Beihang University, Beijing 100191, China\\
Email: huasenwu@gmail.com}
\IEEEauthorblockA{\IEEEauthorrefmark{2}Department of Computer Science, University of California, Davis, CA 95616, USA}
\IEEEauthorblockA{\IEEEauthorrefmark{3}Mallard Creek Networks, 11452
Mallard Creek Trail, Fairfax, VA 22033, USA}
\IEEEauthorblockA{\IEEEauthorrefmark{4}Department of Electrical and
Computer Engineering, University of Maryland, College Park, MD 20742, USA}}

\maketitle

\begin{abstract}
%Event-driven Machine-to-Machine (M2M) communication is characterized by highly bursty traffic, where a large number of devices activate within a short period of time. In such cases, response time depends primarily on random access delay. After analyzing the limit of traditional access control strategies, this paper proposes a Fast Adaptive S-ALOHA (FASA) scheme for M2M communication systems with bursty traffic. In FASA, the statistics of consecutive idle and collision slots are used to accelerate the tracking process of network status. We use drift analysis to design the scheme and to analyze its asymptotic behavior. We show that with FASA, the estimate of the number of backlogged devices is drawn quickly to the true value and the access delay can be reduced. Simulation results show that for a large number of active devices, the access delay of the proposed FASA scheme is rather close to the theoretical optimum value, much better than that of traditional additive schemes such as PB-ALOHA. Compared to multiplicative schemes, e.g., Q$^+$-Algorithm, the proposed FASA scheme has better stability performance under heavy load in addition to slightly better delay performance. Hence, the proposed FASA scheme is more applicable for random access control of M2M communications.

Machine-to-Machine (M2M) communication is now playing a market-changing role in a wide range of business world. However, in event-driven M2M communications, a large number of devices activate within a short period of time, which in turn causes high radio congestions and severe access delay. To address this issue, we propose a Fast Adaptive S-ALOHA (FASA) scheme for M2M communication systems with bursty traffic. The statistics of consecutive idle and collision slots, rather than the observation in a single slot, are used in FASA to accelerate the tracking process of network status. Furthermore, the fast convergence property of FASA is guaranteed by using drift analysis. Simulation results demonstrate that the proposed FASA scheme achieves near-optimal performance in reducing access delay, which outperforms that of traditional additive schemes such as PB-ALOHA. Moreover, compared to multiplicative schemes, FASA shows its robustness even under heavy traffic load in addition to better delay performance.
\end{abstract}

%%%%%%%%%%
\section{Introduction}\label{sec:introd}
%%%%%%%%%%
Machine-to-Machine (M2M) communication or Machine-Type Communication (MTC) is expected to be one of the major driving forces of cellular networks, as its demand is increasing greatly in recent years \cite{3GPP2011TS22368,Cisco2011TR}. Behind the proliferation of M2M communication, the congestion problems in M2M communication become a big concern. The reason is that the device density of M2M communication is much higher than that in traditional Human-to-Human (H2H) communication \cite{3GPP2011TS22368}. What's worse, in event-driven M2M applications, many devices may be triggered almost simultaneously and attempt to access the base
station (BS) through the Random Access Channel (RACH) \cite{Bertrand2009RA}. Such high burstiness can result in congestion
and increase response time, which motivates our research.

In literature, several strategies have been proposed for avoiding radio congestions in M2M communication \cite{Lien&Chen2011CL,Kim2011CCIS,3GPP2010R2_104662,3GPP2010R2_100182,Wang2010ICWITS}. Among these strategies, Slotted-ALOHA (S-ALOHA) type policies, e.g., access class barring (ACB) based schemes in \cite{3GPP2010R2_104662,3GPP2010R2_100182}, are applied in 3GPP for random access control of M2M devices. In these schemes, it is left to users to
decide the operation parameters such as transmission probability to stabilize and optimize the system. Two typical classes of schemes, additive and multiplicative schemes, have been proposed for stabilizing the S-ALOHA system. Historical outcomes are applied in these schemes to estimate the network status and optimize the access probability. However, as discussed in more detail later, traditional additive schemes, such as Pseudo Bayesian ALOHA (PB-ALOHA) \cite{Rivest1987PBALOHA}, estimate the number of backlogged devices based on the observation of the previous slot and cannot adjust the transmission probability in time under highly busty traffic, which results in large access delay. The delay can be shortened in multiplicative schemes \cite{Hajek1982ITAC}, e.g., Q-Algorithm \cite{ISO18000_6} and its enhanced version Q$^+$-Algorithm \cite{Lee2007QPlus}. The reason is that the estimate has exponential increment in consecutive collision slots and exponential decrement in idle slots, which means multiplicative schemes can track the network status in a short period. However, the throughput suffers in these schemes due to the fluctuations in the estimation \cite{Hajek1982ITAC}. More recently, Adaptive Traffic Load Slotted Multiple Access Collision Avoidance (ATL S-MACA) mechanism in \cite{Wang2010ICWITS} uses packet sensing and adaptive method to improve the access performance under heavy traffic load. But the scheme is designed for M2M communications with Poisson traffic and is not suitable for event-driven M2M applications.

In this paper, we propose a Fast Adaptive S-ALOHA (FASA) scheme for access control of event-driven M2M communications. In order to deal with congestions resulted from the burstiness, we collect access results in the past slots, in particular,
consecutive idles or collisions, and apply them to track the network
status. Furthermore, using drift analysis, we carefully design the parameters in the scheme such that the transmission probability can converge quickly to the optimal value. With numerical simulations, we show that the proposed FASA scheme can achieve the near-optimal performance in reducing the access delay, as well as robust performance under all traffic loads less than $e^{-1}$, which is the maximum throughput of S-ALOHA system.

The remainder of the paper is organized as follows. In
Section~\ref{sec:sys_model}, we present the system model, including
the bursty traffic model for event-driven M2M communications. In
Section~\ref{sec:algorithm}, after analyzing the limit of
traditional fixed step-size adaptive policies, we propose the FASA scheme and design it based on drift analysis. In
Section~\ref{sec:sim_res}, simulation results are presented to
evaluate the performance of the proposed scheme, compared with the
theoretical optimal scheme, PB-ALOHA, and Q$^+$-Algorithm. In
Section~\ref{sec:conclusion}, we discuss future work and conclude
the paper.

%%%%%%%%%%
\section{System Model}\label{sec:sys_model}
%%%%%%%%%%
In this paper, we consider S-ALOHA based random access control for
bursty M2M communication traffic. The system consists of a BS and a
large number of M2M devices, where devices with data attempt to
access the BS through a single RACH. The time is divided into time
slots, each of which is long enough to transmit a request packet.
Deferred first transmission (DFT) mode \cite{Rivest1987PBALOHA} is
assumed, in which a device with a new request packet immediately
goes to backlogged state. In slot $t$, all backlogged devices
transmit packets with probability $p_t$, which is broadcasted by the
BS at the beginning of the slot. Moreover, an ideal collision
channel is assumed, where the transmitted packet will be
successfully received by the BS when no other packets are being
transmitted in the same slot. Let $Z_t$ denote the access result in
slot $t$, and $Z_t = 0$, 1, or $c$ depending on whether zero, one,
or more than one request packets are transmitted on RACH. At the end
of slot $t$, the BS decides the transmission probability for next
slot based on the sequence $\{Z_0,Z_1, \ldots, Z_t\}$, i.e.,
\begin{equation} \label{eq:decide_tran_prob}
p_{t+1} = \Pi_t(Z_0, Z_1, \ldots, Z_t).
\end{equation}
The objective of the BS is to maximize the throughput and minimize
the access delay. It is well known that, when $N_t \geq 1$ in slot $t$, where $N_t$ is the number of backlogged devices in slot $t$, using transmission probability $p_t = 1 / N_t$ maximizes the throughput of the S-ALOHA system. However, the BS does not know $N_t$
and has to obtain its estimate $\hat{N}_t$ based on the access
results in the past.

For the traffic model, we focus on event-driven M2M applications,
i.e., when an event is detected, a random and large number of M2M
devices become active almost simultaneously and attempt to access
the BS on the RACH. We call this an active stage. To capture the
burstiness of M2M traffic, instead of assuming Poisson arrival
process, we assume that when an event is detected, $N$ devices are
triggered in a short duration $T$. As suggested in
\cite{3GPP2010R2_104662}, the active time follows the beta
distribution, of which the probability density function is
\begin{equation} \label{eq:active_dist}
f(x) = \frac{x^{\alpha-1}(T-x)^{\beta-1}}{T^{\alpha + \beta
-1}B(\alpha, \beta)}, ~~x \in [0,T],
\end{equation}
where $B(\alpha, \beta)$ is the beta function, and \cite{3GPP2010R2_104662} suggests that the values $\alpha = 3$ and
$\beta = 4$ give the best fit. The number of active devices $N$ and the
active duration $T$ are both random variables and no prior knowledge
about them is assumed at the BS. For the sake of simplicity, a
backlogged device will not generate any new requests since the new
coming data can be transmitted as long as the device accesses the BS
successfully. On the other hand, all the backlogged devices will
keep retransmitting until their request packets are successfully
received by the BS.

\section{Fast Adaptive S-ALOHA}\label{sec:algorithm}
%%%%%%%%%%
%In a S-ALOHA system, BS decides the transmission probability $p_t$
%based on the access results in the past slots. Let $Z_t$ denote the
%access result in slot $t$, and $Z_t = 0$, 1 or $c$ depending on
%whether zero, one, or more than one request packets are transmitted
%on RACH. At the end of slot $t$, the BS calculates the transmission
%probability based on the sequence $\{Z_1, Z_2, \ldots, Z_t\}$, i.e.,
%\begin{equation} \label{eq:}
%p_{t+1} = \Pi(Z_1, Z_2, \ldots, Z_t).
%\end{equation}
%
%It is easy to show that using a transmission probability $p_{t} =
%1/N_t$ in slot $t$ will maximize the throughput of the S-ALOHA
%system, where $N_t$ is the number of backlogged devices in slot $t$.
%However, BS does not know the number of backlogged devices and many
%schemes have been proposed to estimate it
%\cite{Hajek1982ITAC,Rivest1987PBALOHA}.

The estimation of the number of backlogged devices plays an important part in stabilizing and optimizing the S-ALOHA system. In
this section, using drift analysis, we first examine the limit of traditional fixed step-size estimation schemes. Then, we propose and analyze a fast adaptive scheme, referred to as Fast Adaptive S-ALOHA.

\subsection{Drift Analysis of Fixed Step-size Estimation Schemes}
Many additive schemes with fixed step-size have been proposed to estimate the
number of backlogged devices. A unified framework of these schemes
is proposed and studied by Kelly in \cite{Kelly1985StochModels},
where the estimate $\hat{N}_{t}$ is updated by the recursion
\begin{equation} \label{eq:est_framework}
\hat{N}_{t+1} = \max\{1, \hat{N}_t + a_0 I[Z_t = 0] + a_1 I[Z_t = 1]
+ a_cI[Z_t = c]\},
\end{equation}
where $a_0$, $a_1$, and $a_c$ are constants and $I[A]$ is the
indicator function of event $A$.

With the estimation, the BS sets the transmission probability $p_t =
1/\hat{N}_t$ for all backlogged devices, and thus the offered load
$\rho = N_t p_t = N_t/\hat{N}_t$, representing the average number of
devices attempting to access the channel. To stabilize and optimize
the S-ALOHA system, $\hat{N}_t$ needs to track the actual number of
backlogged devices $N_t$, especially when $N_t$ is large. When $N_t
= n$ and $\hat{N}_t = \hat{n}$, the estimation drift can be
calculated as \cite{Kelly1985StochModels}
\begin{eqnarray} \label{eq:est_drift}
 &&E[\hat{N}_{t+1}-\hat{N}_t|N_t = n, \hat{N}_t=
\hat{n}]~~~~~~~~~~~~\nonumber\\
 &=&(a_0-a_c)\left(1-\frac{1}{\hat{n}}\right)^n +
(a_1-a_c)\frac{n}{\hat{n}}\left(1-\frac{1}{\hat{n}}\right)^{n-1}
+a_c \nonumber\\
& \to &(a_0-a_c)e^{-\rho} + (a_1-a_c)\rho e^{-\rho} + a_c \overset
{\rm def}{=} \Delta(\rho),
\end{eqnarray}
as $n \to \infty$, with $n/\hat{n} = \rho$ fixed.

%For PB-ALOHA \cite{Rivest1987PBALOHA}, $a_0 = a_1 = -1 +
%\hat{\lambda}_t$, and $a_c = (e-2)^{-1}+ \hat{\lambda}_t$, where
%$\hat{\lambda}_t$ is the estimate of arrival rate. It is suggested
%in \cite{Tsitsiklis1987StabAnal} that using $\hat{\lambda}_t =
%e^{-1}$ stabilizes the S-ALOHA system for Poisson arrival process
%with mean $\lambda < e^{-1}$, since the drift of PB-ALOHA satisfies
%$\Delta_{PB-ALOHA}(\rho) < 0$ if $\rho <1$ and
%$\Delta_{PB-ALOHA}(\rho) > 0$ if $\rho >1$.

By properly choosing the parameters $a_i$ ($i = 0, 1, c$) such that $\Delta(\rho) < 0$ if $\rho <1$ and
$\Delta(\rho) > 0$ if $\rho >1$, the estimate $\hat{N}_t$ will drift towards the true value and thus the S-ALOHA system can be stabilized. However, these fixed step-size schemes are not suitable for systems with bursty traffic. When the estimate $\hat{N}_t$ deviates far away from the true value $N_t$, we have $\lim_{\rho \to 0} \Delta(\rho) = a_0$ and $\lim_{\rho \to \infty} \Delta(\rho) = a_c$. These limits indicate that the drift tends to be a constant even when the deviation is large, which could result in a large tracking time. Thus, it is necessary to design fast estimation schemes for event-driven M2M communication.

\subsection{Design of FASA}
As analyzed in the previous subsection, fixed step-size estimation
schemes such as PB-ALOHA may not be sufficient to adapt in a timely
manner for systems with bursty traffic because it uses a
constant step-size even when the estimate is far away from the true
value. We note that in addition to the access result in the previous
slot, the access results in several consecutive slots will be
helpful for improving the estimation as they may reveal additional
information about the true value. Intuitively, collisions in several
consecutive slots are likely caused by a significant
underestimation, i.e., $\hat{N}_t \ll N_t$, and the BS should
aggressively increase its estimate. Similarly, several consecutive
idle slots may indicate that the estimate $\hat{N}_t \gg N_t$, and
it should be reduced aggressively.

Motivated by this intuition, we propose a FASA scheme that updates $\hat{N}_t$ as follows
\begin{equation} \label{eq:} \hat{N}_{t+1} =
\left\{\begin{aligned}
&\max\{1, \hat{N}_{t} -1 - h_0(\nu)(K_{0,t})^{\nu}\},&{\rm if}~~Z_t = 0\\
&\hat{N}_{t},&{\rm if}~~Z_t = 1\\
&\hat{N}_{t} + \frac{1}{e-2}  + h_c(\nu)(K_{c,t})^{\nu},&{\rm
if}~~Z_t = c
\end{aligned}\right.
\end{equation}
where $K_{0,t}$ and $K_{c,t}$ are the numbers of consecutive idle
and collision slots up to slot $t$, respectively; $\nu > 0$ is the
parameter that controls the adjusting speed; $h_0(\nu)$ and
$h_c(\nu)$ are functions of $\nu$ that guarantee the tracking
criterions \cite{Kelly1985StochModels}. We are mostly interested in
those cases where the number of backlogged devices is large. Hence we
will approximate $\max\{1,x\}$ as $x$ for notational simplicity in
the analysis later.

Next, we design $h_0(\nu)$ and $h_c(\nu)$ by drift analysis.
Consider tracking a fixed number of backlogged devices, i.e., $N_t =
n$ is constant for all $t \in \mathbb{Z}_+:=\{0,1,2,\ldots\}$.
Assume that in slot $t$, the estimate $\hat{N}_t = \hat{n}$, and
thus the offered load $\rho = n/\hat{n}$. When $n$ is large, the
drift of estimation for FASA can be calculated as
\begin{eqnarray} \label{eq:est_drift_fasa}
 &\Delta_{FASA}(\rho) = E[\hat{N}_{t+1}-\hat{N}_t]~~~~~~~~~~~~~~~~~~~~~~~~~~~~~~~\nonumber\\
 &=q_0(\rho)E[\Delta_0(\rho)] + q_1(\rho)E[\Delta_1(\rho)] +
 q_c(\rho)E[\Delta_c(\rho)],
\end{eqnarray}
where $q_0(\rho) = e^{-\rho}$, $q_1(\rho) = \rho e^{-\rho}$, and
$q_c(\rho) = 1 - q_0(\rho) - q_1(\rho)$ are the probabilities of an
{\it{idle}}, {\it{success}}, and {\it{collision}} slot,
respectively; $\Delta_i(\rho)$ ($i = 0,1,c$) is the change in
$\hat{N}_t$ resulting from the corresponding update.

Obviously, $\Delta_1(\rho) = 0$ since the estimated number remains
unchanged when a packet is successfully transmitted in a slot.
However, $\Delta_0(\rho)$ and $\Delta_c(\rho)$ depend on the
distribution of $K_{0,t}$ and $K_{c,t}$, respectively. For example,
\begin{eqnarray} \label{eq:}
E[\Delta_0(\rho)] = \sum_{k_0 = 1}^{t+1}E[\Delta_0(\rho)|K_{0,t} =
k_0]Pr(K_{0,t} = k_0).\nonumber
\end{eqnarray}

Unfortunately, it is difficult to obtain the distribution of
$K_{0,t}$ and $K_{c,t}$ directly, and thus we resort to
approximation in order to make the problem tractable. It is easy to
see that the update of $\hat{N}_t$ is mostly affected by the access
results in the past $s$ slots $\{Z_{t-s},Z_{t-s+1},...,Z_{t-1}\}$,
where $s$ is the number of consecutive idles or collisions
immediately proceeding $t$, or $s = 1$ if $Z_{t-1}=1$. The update
step is reset whenever $Z_{t-1} \neq Z_t$ or $Z_t=1$. For
tractability, we approximate the estimates $\hat{N}_{t'}$ $(t' =
0,1,\ldots,t-1)$ as $\hat{N}_t$. Thus, the access outcomes
$\{Z_0,Z_1, ...,Z_t\}$ are independent and identically distributed
(i.i.d.) random variables, of which $q_0(\rho)$, $q_1(\rho)$, and
$q_c(\rho)$ are the probabilities of the {\it{idle}},
{\it{success}}, and {\it{collision}} slot, respectively. As we will
see later, this provides a rough bound for the drift and guarantees
the convergence of the algorithm.

First, to calculate the expected drift in an idle slot, suppose that
no packet is transmitted in slot $t$. Then the estimated number will
be reduced by $1 + h_0(\nu)(K_{0,t})^{\nu}$. $K_{0,t} = k_0$ ($k_0
\leq t$) holds when slots $t - k_0 +1, t - k_0 +2, \ldots, t -1$ are
all idle while slot $t - k_0$ is not. Under the approximation that
the access results are i.i.d. in each slot, and the probability that
an idle slot occurs is $q_0(\rho)$, we have
\begin{equation*} \label{eq:} Pr(K_{0,t} = k_0) =
\left\{\begin{aligned}
&q_0^{k_0-1}(\rho)[1-q_0(\rho)],&{\rm if}~~ 1 \leq k_0 \leq t\\
&q_0^{k_0 -1}(\rho),&{\rm if}~~k_0 = t +1.
\end{aligned}\right.\nonumber
\end{equation*}

As $t$ tends to infinity, $q_0^{t}(\rho)$ tends to 0, and $K_{0,t}$
can be approximated by a geometrically distributed random variable
with success probability $1-q_0(\rho)$. Hence,
\begin{align} \label{eq:drift0}
 E[\Delta_0(\rho)]&  \approx \sum_{k_0 = 1}^{\infty}[-1 - h_0(\nu)k_0^\nu]
q_0^{k_0-1}(\rho)[1-q_0(\rho)]\nonumber\\
&= -[1 + h_0(\nu)M(\nu,q_0(\rho))],
\end{align}
where $M(\nu, q_0(\rho))$ is the approximate expectation of
$(K_{0,t})^\nu$ when the offered load is $\rho$, which is given by
\begin{equation} \label{eq:drift0}
M(\nu, q_0(\rho)) = \sum_{k_0 = 1}^{\infty}k_0^\nu
q_0^{k_0-1}(\rho)[1-q_0(\rho)].
\end{equation}

Secondly, we can calculate the drift of the estimation in a collision
slot in a similar fashion as follows:
\begin{align} \label{eq:drift2}
 E[\Delta_c(\rho)] \approx (e-2)^{-1} + h_c(\nu)M(\nu, q_c(\rho)).
\end{align}

From (\ref{eq:drift0}) and (\ref{eq:drift2}), the drift of FASA can
be approximated using
\begin{eqnarray} \label{eq:est_drift_fasa2}
 &\Delta_{FASA}(\rho) \approx -q_0(\rho)[1 + h_0(\nu)M(\nu, q_0(\rho))]\nonumber\\
 & + q_c(\rho)[(e-2)^{-1} + h_c(\nu)M(\nu, q_c(\rho))].
\end{eqnarray}

In order to keep the estimated number $\hat{N}_t$ close to the true
value, it is required that $\Delta_{FASA}(\rho) = 0$ for $\rho = 1$.
Letting $q_0^* = q_0(1) = e^{-1}$ and $q_c^* = q_c(1) = 1-
2e^{-1}$, we expect that
\begin{align} \label{eq:est_drift_fasa_equilium}
 &&\Delta_{FASA}(1) = -h_0(\nu)q_0^*M(\nu, q_0^*)+ h_c(\nu)q_c^*M(\nu,
 q_c^*) = 0.
\end{align}
Therefore, in order to satisfy the condition in
(\ref{eq:est_drift_fasa_equilium}), we can select the following
$h_0(\nu)$ and $h_c(\nu)$:
\begin{equation} \label{eq:norm_function_idle}
h_0(\nu) = \eta [q_0^*M(\nu, q_0^*)]^{-1},
\end{equation}
\begin{equation} \label{eq:norm_function_coll}
h_c(\nu) = \eta [q_c^*M(\nu, q_c^*)]^{-1},
\end{equation}
where $\eta > 0$ is a constant and is another parameter for controlling the tracking speed.

We now interpret the effect of i.i.d. approximation. We focus on the
cases of $\rho \ll 1$ and $\rho \gg 1$, since the evolution of
$\hat{N}_t$ in these cases diverges the most from the i.i.d.
approximation. If $\rho = n/\hat{n} \ll 1$ in slot $t$, it is more
likely that there were consecutive idle slots and the estimate
$\hat{N}_t$ was decreasing in the past slots, suggesting
$\hat{N}_{t'} > \hat{N}_t$ for slot $t' < t$ with high probability.
In this case, the {\it{idle}} probabilities in the past slots would
have been larger than that in slot $t$, while the {\it{collision}}
probabilities would have been smaller than that in slot $t$. Hence, using
$\hat{N}_t$ as an approximation of $\hat{N}_{t'}$ $(t' < t)$
overestimates $E[\Delta_0(\rho)]$ and $E[\Delta_c(\rho)]$, and the drift analysis above gives an upper bound of the drift for $\rho
\ll 1$. Similarly, when $\rho \gg 1$, the drift analysis above gives
a lower bound of the drift. Subsequently, the above analysis using
the i.i.d. approximation roughly bounds the evolution of $\hat{N}_t$
in both directions and guarantees the convergence of $\hat{N}_t$ to
$n$ following Proposition \ref{thm:stable_of_fasa} (Section \ref{subsec:drift_analysis}).

\subsection{Drift Analysis of FASA} \label{subsec:drift_analysis}
The chosen $h_0(\nu)$ and $h_c(\nu)$ guarantee that
$\Delta_{FASA}(1) = 0$ and thus provide a necessary condition for
FASA to track the number of backlogged devices. Furthermore, Proposition \ref{thm:stable_of_fasa}, which can be proved by examining the derivative of $\Delta_{FASA}(\rho)$, shows a desirable property of
FASA. With this property, the BS inclines to decrease $\hat{N}_t$
when $\hat{N}_t > N_t$ and increase $\hat{N}_t$ when $\hat{N}_t <
N_t$. Thus, $\hat{N}_t$ is able to track the number of backlogged
devices $n$.

\begin{prop} \label{thm:stable_of_fasa}
Given that $h_0(\nu)$ and $h_c(\nu)$ are defined in
(\ref{eq:norm_function_idle}) and (\ref{eq:norm_function_coll}),
respectively, the approximate drift of FASA $\Delta_{FASA}(\rho)$ is
a strictly increasing function of $\rho$. In addition,
$\Delta_{FASA}(\rho) < 0$ when $0 < \rho <1$ and
$\Delta_{FASA}(\rho)>0$ when $\rho >1$.
\end{prop}
\proof See Appendix.

In order to understand better the behavior of the scheme, we now present the drift of estimation for $\nu = 1, 2$ and 3. When $\nu \in \mathbb{Z}_+$, $M(\nu, q_i(\rho))$ $(i = 0, c)$ is the
$\nu$th-moment of a geometrically distributed random variable with
success probability $1-q_i(\rho)$, and its closed-form expression
can be obtained. Consequently, with the expressions of $M(\nu, q_i(\rho))$, we can obtain functions $h_0(\nu)$, $h_c(\nu)$, and the drift $\Delta_{FASA}(\rho)$ for $\nu = 1, 2$ and 3. For example, when $\nu = 2$, the drift of FASA can be expressed as
\begin{eqnarray} \label{eq:est_drift_nu2}
\Delta_{FASA}^{(2)}(\rho) =\Delta^{(0)}
+\eta\left[-\frac{(e-1)^2(e^\rho+1)}{(e+1)(e^\rho
-1)^2}\right.\nonumber\\
+\left.\frac{2(e^\rho-\rho-1)(2e^\rho-\rho-1)}{(e-2)(e-1)(\rho
+1)^2}\right],
\end{eqnarray}
where $\Delta^{(0)} = -e^{-\rho}+\frac{1-2e^{-\rho}}{e-2}$ is the
drift of the fixed step-size estimation (\ref{eq:est_framework})
with $a_0 = -1$, $a_1 = 0$ and $a_c = (e-2)^{-1}$.

%When the number of backlogged devices $N_t \ll \hat{N}_t$, and thus
%$\rho \approx 0$, with (\ref{eq:est_drift_nu1}),
%(\ref{eq:est_drift_nu2}), and (\ref{eq:est_drift_nu3}), according to
%L'H\^opital's rule, we have
%\begin{eqnarray} \label{eq:}
%%\Delta_{FASA}^{(1)}(\rho) \sim -\frac{b_1}{\rho},~
%\Delta_{FASA}^{(\nu)}(\rho) \sim -\frac{b_\nu}{\rho^\nu},
%%\Delta_{FASA}^{(3)}(\rho) \sim -\frac{b_3}{\rho^3},
%\end{eqnarray}
%for $\nu = 1,2,3$, where $b_\nu >0$ $(\nu = 1,2,3)$ are constants
%and $f_1(\rho)\sim f_2(\rho)$ means that $f_1(\rho)/ f_2(\rho)$
%tends to one as $\rho \to 0$. Likewise, as $\rho \to \infty$,
%\begin{eqnarray} \label{eq:}
%%\Delta_{FASA}^{(1)}(\rho) \sim \frac{b_1'e^\rho}{\rho},
%\Delta_{FASA}^{(\nu)}(\rho) \sim
%b_\nu'\left(\frac{e^\rho}{\rho}\right)^\nu,
%%\Delta_{FASA}^{(3)}(\rho)\sim  b_3'\left(\frac{e^\rho}{\rho}\right)^3.
%\end{eqnarray}
%for $\nu = 1,2,3$, where $b_\nu' (\nu = 1,2,3)$ are positive
%constants.

\begin{figure}[htbp]
\begin{center}
\includegraphics[angle = 0,width = 0.85\linewidth]{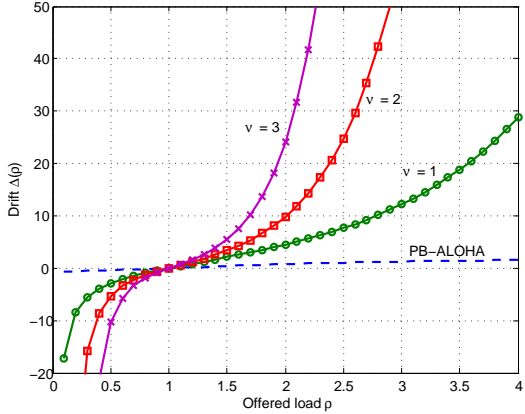}
\caption{Drift of estimation ($\eta = 1$)} \label{fig:drift_of_est}
\end{center}
\vspace{-0.5cm}
\end{figure}

As shown in Fig.~\ref{fig:drift_of_est}, when the estimated number
deviates far away from the actual number of backlogged devices, FASA
adjusts its step-size accordingly, while PB-ALOHA still uses the
same step-size. Therefore, using FASA results in much shorter
adjusting time than PB-ALOHA, and thus improves the performance of
M2M communication systems with bursty traffic. Note that the drifts of multiplicative schemes such as $Q^+$-Algorithm are not illustrated here since they depend on not only the offered load $\rho$ but also the estimate $\hat{N}_t$.

%%%%%%%%%%
\section{Simulation Results}\label{sec:sim_res}
%%%%%%%%%%
In this section we evaluate the performance of the proposed scheme
through simulation. We compare the performance of our FASA scheme,
the ideal policy with perfect knowledge of backlog, PB-ALOHA
\cite{Rivest1987PBALOHA}, and Q$^+$-Algorithm \cite{Lee2007QPlus}.

\subsection{Settings}
First, we choose the parameter values for the adaptive schemes. Since simulations show that the performances of FASA with different $\eta$ and $\nu$ are close, we just present the results for $\eta =1$ and $\nu = 2$ in this paper due to the space limitation. With perfect knowledge of $N_t$, the ideal policy sets transmission
probability at $p_t = 1/N_t$ for $N_t > 0$. Thus, the ideal policy achieves the minimum access delay of S-ALOHA and serves as a benchmark in the comparison. For PB-ALOHA, we use the estimated arrival rate
$\hat{\lambda}_t = e^{-1}$, as suggested in
\cite{Tsitsiklis1987StabAnal}. Q$^+$-Algorithm belongs to the class of
multiplicative schemes which is first proposed by Hajek and van Loon
\cite{Hajek1982ITAC}. In Q$^+$-Algorithm, $\hat{N}_t$ is updated as
follows:
\begin{equation*} \label{eq:est_scheme_qplus}
\hat{N}_{t+1} = \max\{1, [I(Z_t = 0)/\zeta_0 + I(Z_t = 1) +\zeta_c
I(Z_t = c)] \hat{N}_{t}\}, \nonumber
\end{equation*}
where $\zeta_0 = 2^{0.25} \approx 1.1892$ and $\zeta_c = 2^{0.35}
\approx 1.2746$ are suggested in \cite{Lee2007QPlus} for optimal
performance.
%The Q$^+$-Algorithm has a frame-based structure with
%quantized frame length, which usually updates the estimate at the
%end of each frame and sets the frame length to the power of 2. In
%order to obtain a fair comparison, the estimation procedure is
%carried out after each slot and continuous transmission probability
%is applied.

Second, we describe the simulation scenarios. In order to gain more insights into the performance of access schemes
with bursty traffic, we present simulation results for a single
active stage and assume that the intervals between two consecutive
active stages are sufficiently large so that all the devices can
successfully access the BS before the next trigger. During the
active stage, $N$ devices are triggered according to the beta
distribution with parameters $\alpha = 3$, $\beta = 4$
\cite{3GPP2010R2_104662}, and time span $T = 50$ slots, which is the
number of slots in one second when the PRACH period
\cite{Bertrand2009RA} is 20 ms. It is assumed in 3GPP that 30,000
devices can be triggered in 10 s \cite{3GPP2010R2_104662}, thus we
choose $N$ to be 100 to 3,000 per second.

%\subsection{The Effect of $\nu$ and $\eta$}
%Two parameters $\nu$ and $\eta$ are used in the proposed FASA scheme
%to control the adaptive rate of the step-size.We examine the access
%delay under different values of them. As shown in the top figure of
%Fig.~\ref{fig:averdelay_diff_param}, the average access delay is
%rather close for $\nu = 1, 2, 3$ with fixed $\eta = 1$. While the
%delay of the scheme with $\nu = 1$ and $3$ is slightly larger than
%that with $\nu = 2$, they are all close to that with perfect
%information. Similar conclusion can be derived from the lower part
%of the figure for $\eta = 0.5, 1, 2$ with fixed $\nu = 2$, where the
%access delay is quite close and $\eta = 1$ seems to be the best
%option for different numbers of active devices. Again, all delays
%are very close to that with perfect information. According to the
%drift analysis in Section~\ref{sec:algorithm}, with larger $\nu$ and
%$\eta$, the scheme can adjust the estimate more quickly to the
%actual number of backlogged devices, but will result in larger
%variation. Thus, the choice of parameters represents a trade-off
%between the tracking time and estimation accuracy. We suggest $\nu =
%2$ and $\eta = 1$ for the proposed FASA scheme since this setting
%performs slightly better than others in our simulations.
%
%\begin{figure}[htbp]
%\begin{center}
%\includegraphics[angle = 0,width = 0.8\linewidth]{fig/averdelay_diff_param}
%\caption{Effect of $\nu$ and $\eta$}
%\label{fig:averdelay_diff_param}
%\end{center}
%\vspace{-0.5cm}
%\end{figure}

%\subsection{Comparative Study}
\subsection{Results}
In some event-driven M2M
applications, response can be taken with partial messages from the
 detecting devices and not all devices need to report an event. Thus, both the
distribution of access delay and average delay are evaluated to
study the performance of the proposed scheme.

\begin{figure}[htbp]
\begin{center}
\includegraphics[angle = 0,width = 0.85\linewidth]{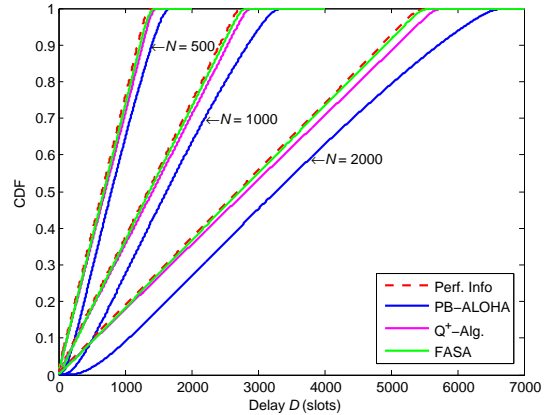}
\caption{Cumulative distribution function of access delay}
\label{fig:delay_cdf}
\end{center}
\vspace{-0.5cm}
\end{figure}

Fig.~\ref{fig:delay_cdf} shows the cumulative distribution function
of access delay for different schemes. From
this figure we can see that the performance of the proposed FASA
scheme is close to the benchmark with perfect knowledge. For
PB-ALOHA, it takes a long time to track the number of backlogged
devices and few devices can access successfully during this period.
For instance, the 10\% delay, which is the access delay achieved by
10\% of the active devices, is much larger than other schemes. For
example, when $N = 500$, the 10\% delay of FASA is about 280 slots
while it is 520 slots for PB-ALOHA. With multiplicative increment,
the Q$^+$-Algorithm can track the number of backlogs in a short time
because of the exponential increment due to the consecutive
collision slots. However, it takes longer for all the devices to
access the channel under Q$^+$-Algorithm than under FASA due to the
large estimation fluctuation in Q$^+$-Algorithm.

\begin{figure}[htbp]
\begin{center}
\includegraphics[angle = 0,width = 0.85\linewidth]{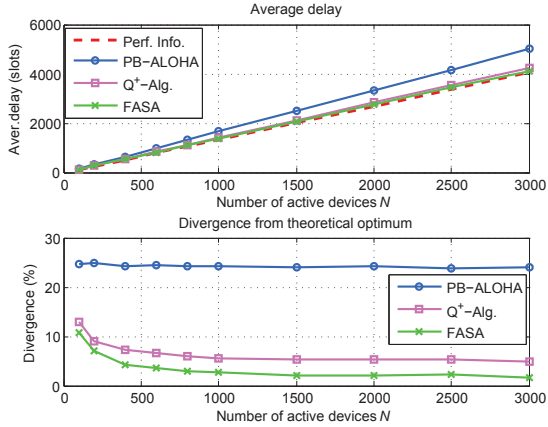}
\caption{Average delays for different schemes}
\label{fig:averdelay_diff_scheme}
\end{center}
\vspace{-0.5cm}
\end{figure}

Fig.~\ref{fig:averdelay_diff_scheme} compares the average delay of
the access schemes. The top figure indicates that the average delay
increases almost linearly as the number of active devices increases.
To quantify the divergence from the theoretical optimum performance,
we define the normalized divergence as $e(D) = \frac{D-D^*}{D^*} \times 100\%$,
%\begin{equation} \label{eq:}
%e(D) = \frac{D-D^*}{D^*},\nonumber
%\end{equation}
where $D$ is the average delay of a particular scheme and $D^*$ is
the theoretical optimum delay with perfect information. The bottom
part of Fig.~\ref{fig:averdelay_diff_scheme} shows the divergence of
PB-ALOHA, Q$^+$-Algorithm, and FASA. The delay of PB-ALOHA scheme is
larger than the optimum value by about $22\%$. The divergences of
Q$^+$-Algorithm and FASA are both much less than that of PB-ALOHA.
As the number of the active devices increases, the effect of
estimation fluctuation becomes small and the performance gets close
to the optimum value. For large $N$, the divergence is about $5\%$
for Q$^+$-Algorithm and $2\%$ for FASA, which indicates that FASA
scheme performs slightly better than Q$^+$-Algorithm.

As shown in Figs.~\ref{fig:delay_cdf} and
\ref{fig:averdelay_diff_scheme}, for a single active stage, the
performance of FASA and Q$^+$-Algorithm is close, with the proposed
FASA scheme performing slightly better. However, when considering the long
term performance, as discussed in \cite{Hajek1982ITAC}, for given
$\zeta_0$ and $\zeta_1$, the estimated value $\hat{N}_t$ continues
to fluctuate when it gets close to $N_t$ and the stable throughput
is less than some maximum value $\Theta_{\max} < e^{-1}$. We verify
this property through simulations for Poisson arrival process with
mean $\lambda$. The results show that, for PB-ALOHA and FASA, the
number of backlogged devices is finite though the average value
becomes large as $\lambda$ becomes close to $e^{-1}$. For
Q$^+$-Algorithm with the given parameter values, however, when $\lambda$ is
larger than about $0.36$, the number of backlogged devices grows
unbounded, indicating that the algorithm is unstable when $\lambda >
0.36$.

%%%%%%%%%%
\section{Conclusion and Future Work}\label{sec:conclusion}
%%%%%%%%%%
This paper proposes a fast adaptive S-ALOHA scheme, called FASA, for
event-driven M2M communications. An approximate drift analysis and
simulation results show that using FASA, a BS can track the number of
backlogged devices more quickly. That is a main advantage compared to fixed step-size
additive schemes, e.g., PB-ALOHA. Compared to multiplicative
schemes, simulation results show that the proposed FASA scheme has better
stability performance under heavy load in addition to slightly
better delay performance. Currently, the analysis presented in this
paper is based on the approximation with fixed number of backlogged
devices. A rigorous analysis about the properties of the proposed
scheme with general arrival traffic, e.g., mixed bursty and Poisson
traffic, is left as our future work.

%%%%%%%%%%
\section*{Acknowledgement}\label{sec:}
%%%%%%%%%%
%This work was supported in part by the National Basic Research
%Program of China (973 Program, Grant No. 2010CB731803) and the
%National Natural Science Foundation of China (Grant No. 60921001).
This work was supported by the National Basic Research Program of China (973 Program, Grant No. 2010CB731803).

%%%%%%%%%%
\appendices
%%%%%%%%%%

%%%%%%%%%%
\section*{Appendix: Proof of Proposition \ref{thm:stable_of_fasa}}\label{app:proof_of_stable_of_fasa}
%%%%%%%%%%
The proposition can be proved by calculating the derivative of $\Delta_{FASA}(\rho)$.

For a given value of $\nu$, let
\begin{equation} \label{eq:}
g_0(\rho) = |q_0(\rho)E[\Delta_0(\rho)]| = q_0(\rho)[1 +
h_0(\nu)M(\nu, q_0(\rho))],\nonumber
\end{equation}
and
%\begin{align} \label{eq:}
%g_c(\rho) &=  |q_c(\rho)E[\Delta_c(\rho)]| \nonumber\\
%&=  q_c(\rho)[(e-2)^{-1} + h_c(\nu)M(\nu, q_c(\rho))].\nonumber
%\end{align}
\begin{equation} \label{eq:}
g_c(\rho) =  |q_c(\rho)E[\Delta_c(\rho)]| =  q_c(\rho)[\frac{1}{e-2} + h_c(\nu)M(\nu, q_c(\rho))].\nonumber
\end{equation}

Then $g_0(1) = g_c(1) = e^{-1} + \eta$ and $\Delta_{FASA}(1) =
-g_0(1) + g_c(1) = 0$. Next, we claim that, for given $\nu >0$,
$M(\nu, q) = \sum_{k = 1}^\infty k^\nu q^{k-1}(1-q)$ is an
increasing function of $q$ ($0 < q < 1$), because
\begin{eqnarray} \label{eq:}
 \frac{\partial M}{\partial q}&=&\sum_{k = 1}^\infty k^\nu q^{k-2}[k(1-q)-1] \nonumber\\
& = & \sum_{k = 1}^{k^*} k^\nu q^{k-2}[k(1-q)-1]\nonumber\\
&& + \sum_{k = k^*+1}^{\infty} k^\nu q^{k-2}[k(1-q)-1]\nonumber \\
&>&(k^*)^\nu \sum_{k = 1}^{\infty} q^{k-2}[k(1-q)-1] = 0,
\end{eqnarray}
where $k^* = \lfloor \frac{1}{1-q} \rfloor$ is the largest integer
not greater than $\frac{1}{1-q}$, and thus $k^\nu \leq (k^*)^\nu$ if
$1\leq k \leq k^*$ and $k^\nu > (k^*)^\nu$ if $k > k^*$. In
addition, the idle probability $q_0(\rho) = e^{-\rho}$ is
nonnegative and strictly decreasing in $\rho$. Hence, $M(\nu,
q_0(\rho))$ is strictly decreasing in $\rho$ and $g_0(\rho)$ is a
strictly decreasing function of $\rho$. On the other hand, since
$q_c(\rho) = 1-e^{-\rho} -\rho e^{-\rho}$ is nonnegative and
strictly increasing in $\rho$, we can show similarly that
$g_c(\rho)$ is a strictly increasing function of $\rho$. Thus,
$\Delta_{FASA}(\rho) = -g_0(\rho) + g_c(\rho)$ is a strictly
increasing function of $\rho$. Consequently, $\Delta_{FASA}(\rho) <
\Delta_{FASA}(1) =0$ when $0 < \rho < 1$ and $\Delta_{FASA}(\rho)
> \Delta_{FASA}(1) =0$ when $\rho
> 1$.

\bibliography{FA_ALOHA}

\end{document}